\documentclass[twocolumn]{aastex631}

\usepackage{amsmath}
\usepackage{commath}
\usepackage{xcolor}
\usepackage{CJK}

\newcommand{\blazar}{J0410--0139}

\def\mgii     {\ensuremath{\text{Mg\,\textsc{ii}}}}

\def\cii     {\ensuremath{\text{[C\,\textsc{ii]}}}}

\newcommand{\kms}{{\rm km\,s}\ensuremath{^{-1}}}

\begin{document}

\title{[CII] properties and Far-Infrared variability of a $z=7$ blazar}

\correspondingauthor{Eduardo Ba\~nados}
\email{banados@mpia.de}
\author[0000-0002-2931-7824]{Eduardo Ba{\~n}ados}
\affiliation{Max-Planck-Institut f\"ur Astronomie, K\"onigstuhl 17, D-69117, Heidelberg, Germany}

\author[0000-0002-7220-397X]{Yana Khusanova}
\affiliation{Max-Planck-Institut f\"ur Astronomie, K\"onigstuhl 17, D-69117, Heidelberg, Germany}

\author[0000-0002-2662-8803]{Roberto Decarli}
\affiliation{INAF --- Osservatorio di Astrofisica e Scienza dello Spazio, via Gobetti 93/3, I-40129, Bologna, Italy}

\author[0000-0003-3168-5922]{Emmanuel Momjian}
\affiliation{National Radio Astronomy Observatory, P.O. Box O, Socorro, NM 87801, USA}

\author[0000-0003-4793-7880]{Fabian Walter}
\affiliation{Max-Planck-Institut f\"ur Astronomie, K\"onigstuhl 17, D-69117, Heidelberg, Germany}

\author[0000-0002-7898-7664]{Thomas Connor}
\affiliation{Center for Astrophysics $\vert$\ Harvard\ \&\ Smithsonian, 60 Garden St., Cambridge, MA 02138, USA}

\author[0000-0001-6647-3861]{Christopher L. Carilli}
\affiliation{National Radio Astronomy Observatory, P.O. Box O, Socorro, NM 87801, USA}

\author[0000-0002-5941-5214]{Chiara Mazzucchelli}
\affiliation{Instituto de Estudios Astrof\'{\i}sicos, Facultad de Ingenier\'{\i}a y Ciencias, Universidad Diego Portales, Avenida Ejercito Libertador 441, Santiago, Chile.}

\author[0000-0003-2349-9310]{Sof\'ia Rojas-Ruiz}
\affiliation{Department of Physics and Astronomy, University of California, Los Angeles, 430 Portola Plaza, Los Angeles, CA 90095, USA}

\author[0000-0001-9024-8322]{Bram P. Venemans}
\affiliation{Leiden Observatory, Leiden University, Niels Bohrweg 2, 2333 CA Leiden, Netherlands}



\begin{abstract}

We present millimeter observations of the host galaxy of the most distant blazar known, VLASS~J041009.05$-$013919.88 (hereafter J0410--0139) at $z=7$, using ALMA and NOEMA observations. The ALMA data reveal a  $(2.02\pm 0.36)\times 10^{42}$          erg\,s$^{-1}$ \cii\ 158\,$\mu$m emission line at $z=6.9964$ 
with a \cii-inferred star-formation rate of $58\pm 9\,M_\odot\,$yr$^{-1}$. 
We estimate a dynamical mass of $M_{\rm dyn,\cii}=(4.6\pm 2.0)\times 10^9\,M_\odot$, implying a black hole mass to host a dynamical mass ratio of $0.15^{+0.08}_{-0.05}$. 
The 238\,GHz continuum (rest-frame IR) decreased by $\sim$$33\%$ from the NOEMA to the ALMA observations taken $\sim$10 months apart.  The VLA 3--10\,GHz radio flux densities showed a $\sim$$37\%$ decrease in a similar time frame, suggesting a causal connection. At face value, \blazar\ would have the lowest \cii-to-IR luminosity ratio of a $z>5.7$ quasar reported to date ($\sim$$10^{-4}$). However, if only $<$$20\%$ of the measured IR luminosity were due to thermal emission from dust, the \cii-to-IR luminosity ratio would be typical of (U)LIRGS, and the star formation rates derived from \cii\ and IR luminosities would be consistent.  
These results provide further evidence that synchrotron emission significantly contributes to the observed rest-frame IR emission of \blazar, similar to what has been reported in some radio-loud AGN at $z<1$.
\end{abstract}

\keywords{cosmology: observations -- quasars: emission lines  -- quasars: general}


\section{Introduction} \label{sec:intro}

Relativist jets have been well-established as a means for quasars to impact the properties of their host galaxies \citep{hardcastle2020,harrison2024}. Associated with radio-loud quasars, these jets can uplift large amounts of gas, potentially promoting increased star formation in their hosts \citep[e.g.,][]{bieri2015}, or quenching star formation \citep[e.g.,][]{comerford2020}.

To date, the cold dust and gas of only seven galaxies hosting radio-loud quasars at $5.8\!<\!z\!<\!6.9$ have been investigated with the Atacama Large Millimeter/submillimeter Array (ALMA) and NOrthern Extended Millimeter Array (NOEMA), and none beyond $z>6.9$ \citep{rojas2021,khusanova2022,banados2023}. 
 Although the statistics are still limited, their \cii- and IR-derived star-formation properties are consistent with the ones reported for the much more common \citep[e.g.,][]{keller2024}---and thus more studied---radio-quiet quasars (c.f., \citealt{decarli2018,venemans2020}). If galaxy properties are consistent between quasar hosts with and without jets at this epoch, then either the jets are not significantly impacting their galaxies or jets are common but sporadic \citep[e.g.,][]{connor2024}. Fundamental to these arguments is the assumption that the \cii\ and underlying continuum emissions are linked only to star formation and are not associated with the active galactic nucleus (AGN). 
 However, this assumption might be flawed. 

\cite{rojas2021}, studying the $z=5.83$ radio-quasar P352–-15, found that a modified thermal black body model consistent with the quasar host's 290\,GHz flux under-predicted the observed 100\,GHz flux. The interpretation of this discrepancy is that, at 100\,GHz, the flux is being enhanced by contributions from the synchrotron emission from the jets observed at lower frequencies. \cite{khusanova2022}, as part of a study of four $z>6$ quasars with NOEMA, further considered this possibility and described multiple ways that high-redshift radio-loud AGN might impact their observed \cii\ emission, biasing estimates of star formation rate (SFR). Recently, \cite{li2024} studied several CO transitions of the $z=6.18$ radio-quasar J1429+5447, finding the lowest CO excitation among all published $z\sim 6$ quasars with similar data (all the others are radio-quiet).  
 
  Here, we report NOEMA and ALMA observations of a newly discovered radio-quasar at $z_{\rm \mgii}=6.995\pm 0.001$, VLASS~J041009.05$-$013919.88 (hereafter J0410--0139), powered by a black hole with the mass of $6.9^{+0.5}_{-0.4} \times 10^8$\,$M_\odot$ \citep{banados2024a}. Based on its rapid radio variability, X-ray properties, and compact radio emission on pc scales,  \cite{banados2024a} suggests that \blazar\ is a blazar. 
  The new data---separated by 35 days in the quasar's rest frame and matched with closely-timed archival 1--10 GHz radio observations---therefore enable a look at the properties of jet-affected high-redshift quasar host galaxy that accounts for the contributions of the jet emission itself. 
  
  Throughout this Letter, we adopt a standard flat $\Lambda$CDM cosmology with  $H_0 = 70 \,\mbox{km\,s}^{-1}$\,Mpc$^{-1}$, \mbox{$\Omega_M = 0.30$}.

\section{Observations} \label{sec:observations}

\subsection{NOEMA}

To characterize the host galaxy of \blazar, we obtained NOEMA observations targeting the \cii\ emission line ($\nu_{rest}=1900.5369$ GHz) and its underlying continuum emission. The observations were carried out on 16 November 2021 with ten antennae in band 3. The total on-source time was 5.2 hours under good weather conditions with low precipitable water vapor (PWV$\sim1$ mm). We used \texttt{clic} package for calibration, part of the GILDAS software. 
We used LKHA101 for flux density calibration, 3C84 for bandpass calibration, and 0420-014 for phase and amplitude calibrations. 
We produced an image cube with 50 km~s$^{-1}$ resolution using natural weighting and HOGBOM cleaning down to the $2\sigma$ level.  
The data cover the $\sim220-243$\,GHz range and the synthesized beam size is $\sim0\farcs8\times1\farcs6$ with a position angle PA=$9.2^\circ$. 

We produced an image of continuum emission at 224\,GHz by averaging all the channels in the lower side band (LSB). Similarly, we created a 239\,GHz continuum emission image from the upper side band (USB), excluding the channels within $\pm1000$ km~s$^{-1}$ from the expected frequency of the \cii\ line. The continuum emission is well detected in both the LSB and USB images with signal-to-noise ratios S/N$>20$ (Fig.\,\ref{fig:alma_noema}). In both images, the continuum source is unresolved. 
The coordinates of the peak in the 224\,GHz continuum emission are  RA$=$04:10:09.05; DEC$=$$-$01:39:19.90. For the  239\,GHz continuum peak emission, the coordinates are RA$=$04:10:09.05; DEC$=$$-$01:39:19.98. 
These continuum locations are consistent with the quasar's optical position (see Fig.~\ref{fig:alma_noema}). 
We extracted spectra centered on the brightest pixel of the USB continuum map, and no \cii\ emission line is detected from the brightest pixel or a $1\farcs5$ radius aperture.

\subsection{ALMA}

We also obtained ALMA observations of \blazar\ in the C-4 configuration, using 45-46 antennas over three passes: On 28 August 2022 (PWV=2.5--3.0 mm), 29 August 2022 (PWV$\sim$2.8\,mm), and 30 August 2022 (PWV$\sim$1.1\,mm). 
The total on-source time is about 2.5 hours. 
The data were reduced using standard procedures with CASA \citep{mcmullin2007}. 
We produced image cubes with 40\,\kms\ resolution using natural weighting and cleaning to the $2\sigma$ level. 
The synthesized beam is  $0\farcs49\times 0\farcs68$ with PA=$-69^\circ$. 

\begin{figure*}[h!]
\centering
\includegraphics[width=0.5\linewidth]{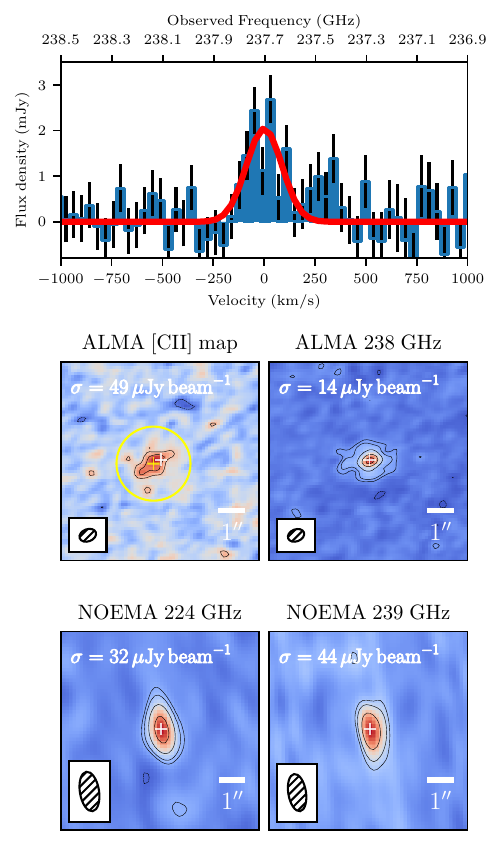}
\caption{\small
\textbf{ALMA and NOEMA observations of \blazar}. \textit{Top:} Continuum-subtracted \cii\ spectrum, extracted from a $1\farcs5$ radius aperture (yellow circle in the next panel). The red line is a Gaussian fit to the data (a second Gaussian can be fitted at $\sim300$ km~s$^{-1}$, but we found this not to be statistically significant; see Section~\ref{sec:ciiprop}). The \cii\ measured properties, luminosity, and star-formation rate are listed in Table~\ref{tab:properties}. The \cii\ line is not detected in the NOEMA data. 
\textit{Middle:} ALMA \cii\ (left) and continuum (right) maps. The beam size for these observations is $0\farcs49\times 0\farcs68$.  The \cii\ map is the average over 247\,\kms\ (1.2$\times$ FWHM of the spectrum shown in the top panel). 
The yellow circle is the $1\farcs5$ radius aperture centered on the \cii\ centroid (yellow cross) used to extract the \cii\ spectrum. 
\textit{Bottom:} NOEMA continuum maps at 224 (left) and 239 (right) GHz, with beam sizes: $0\farcs78\times 1\farcs63$ (left) and $0\farcs72\times 1\farcs53$ (right). The source is unresolved in the NOEMA observations. 
The optical position of the quasar is shown with a white cross on each panel, which is consistent with the peak of the ALMA and NOEMA continuum, but the peak of the \cii\ map is offset by $0\farcs35$ (yellow cross in \cii\ map). 
All contour plots are shown at $\pm [3, 5, 10, 20, 30]\sigma$ (dashed contours represent negative values). The $\sigma$ for each map is at the top left corner of each panel.  
}
\label{fig:alma_noema}
\end{figure*}

Both the continuum and \cii\ line are clearly detected with S/N of 36 and 7, respectively (Fig.~\ref{fig:alma_noema} and Table~\ref{tab:properties}). 
The flux of the \cii\ line is consistent with the noise of the continuum-subtracted NOEMA image, explaining the NOEMA non-detection. 
The peak of the 238\,GHz continuum emission (RA$=$04:10:09.05; DEC$=$$-$01:39:19.91) coincides with the quasar's optical position (see Fig.~\ref{fig:alma_noema}). 
We fit a two-dimensional Gaussian to the 238\,GHz continuum source with the CASA task \texttt{imfit}.  
The source is resolved, with a deconvolved size of $(0.57\pm 0.10)^{\prime\prime}\times (0.68\pm 0.10)^{\prime\prime}$ [$(3.0\pm 0.5)\,\mathrm{kpc}\times (3.5\pm 0.5)\,\mathrm{kpc}$], and a PA of $82^\circ\pm 59^\circ$. 
The centroid of the \cii\ is offset by $0\farcs35$ (1.81\, kpc) from the optical position. We used a curve of growth analysis to determine the optimal aperture radius to extract the spectrum. The cumulative flux density reaches a plateau at $1\farcs5$, independent of whether we use the optical or \cii\ centroid; thus, all the flux density is recovered with this aperture radius. The total \cii\ flux is $0.45\pm 0.08\,$Jy\,km\,s$^{-1}$ (see Table~\ref{tab:properties}). 

\section{Results and discussion}
\label{sec:results}

\subsection{[CII] properties}
\label{sec:ciiprop}

\cii\  is among the brightest emission lines in star-forming galaxies, the main coolant of the neutral interstellar medium, and a tracer of star formation \citep[e.g.,][]{herrera-camus2015}. 
We fit the ALMA \cii\ spectrum shown in Fig.~\ref{fig:alma_noema} with one and two Gaussians. Both fits describe the data well, but neither functional form is preferred according to Bayesian Information Criterion. The fit with two Gaussian reveals a second component shifted by $\sim300$ km~s$^{-1}$. We produced line images centered on each Gaussian by averaging the channels containing the line across 1.2$\times$FWHM, which maximizes the S/N for detecting a Gaussian line (see Appendix A in \citealt{novak2020}). The image of the potential shifted component does not include any channel containing the main component. We did not find any signature above $3.5\sigma$ on the image of the shifted component. We conclude that the shifted component in the spectrum is consistent with noise. Therefore, we use the single Gaussian fit to derive the results presented in Table~\ref{tab:properties} and produce the final \cii\ emission line image averaging channels across 1.2$\times$FWHM (see Fig.~\ref{fig:alma_noema}).

We also show continuum subtracted 40\,\kms\ 
\cii\ channel images in Figure \ref{fig:chmaps30}. 
The emission is highly concentrated in the central channels, although there is some evidence of multiple peaks and different substructures in separate channel images.
While this could indicate a merging system, the evidence is not as compelling as what is found for other systems at similar redshifts \citep[e.g.,][]{banados2019a,neeleman2019,neeleman2021,decarli2017,decarli2019}. Higher S/N data are required to provide conclusive kinematic results.

With our \cii\ data we can make a rough estimate of the dynamical mass of the host galaxy, following \cite{decarli2018}.
We assume a dispersion-dominated system as there is no evidence of rotation in the channel maps of Fig.~\ref{fig:chmaps30}. Thus, the dynamical mass can be expressed as follows:
\begin{equation}
    M_{\rm dyn,\cii} = \frac{3}{2}\,\frac{R_{\cii}\, \sigma^2_{\cii}}{G},
\end{equation}
where $R_{\cii}$ is the major semi-axis of the Gaussian fit of the \cii\ map ($1.75\pm 0.25\,$kpc), $\sigma_\cii$ is the velocity dispersion from the Gaussian fit of the \cii\ line ($88\pm17\,$\kms) and $G$ is the gravitational constant. This results in $M_{\rm dyn,\cii}=(4.6\pm 2.0)\times 10^9\,M_\odot$. 
Assuming that the dynamical mass is comparable to the stellar mass, together with the \cii-derived SFR of  $58\pm 9\,M_\odot\,$yr$^{-1}$, make the host of \blazar\ consistent with a main-sequence galaxy \citep{schreiber2015,algera2023}.

The ratios of the masses of the central supermassive black holes and their host galaxies across cosmic time can provide key insights on galaxy evolution models \citep[e.g.,][]{habouzit2022}. The black hole to galaxy dynamical mass ratio for \blazar\ is $0.15^{+0.08}_{-0.05}$. 
Assuming that the dynamical mass is comparable to stellar or bulge masses, \blazar\ would be $\sim$$2$\,dex above the local black hole-host mass relations \citep{kormendy2013}. This result is consistent with the overmassive supermassive black holes of  $z>6$ quasars identified with ALMA and JWST dynamical mass estimates \citep[e.g.,][]{neeleman2021,marshall2023}.

\subsection{IR Continuum variability}
\label{sec:variability}

The continuum emission from \blazar\ is robustly detected in both NOEMA and ALMA observations. However, the measured flux densities are inconsistent by $\sim 33\%$ (see Table \ref{tab:properties} and Figs.~\ref{fig:alma_noema} and \ref{fig:sed}). 
The ALMA observations were taken 285 days after the NOEMA ones, i.e., about 35 days in the quasar's rest frame. Figure~\ref{fig:sed} also shows the quasi-simultaneous VLA measurements of 2021 and 2022 (taken 39 rest-frame days apart; see Table E1 in \citealt{banados2024a}). 
The VLA flux densities above 3\,GHz show an average decrement of $\sim 37\%$.  
We note that the trend is inverted at observed $\sim$1.5\,GHz, but the emission at frequencies lower than the radio SED peak can be affected by other physical processes, such as free-free absorption or synchrotron self-absorption \citep[e.g.,][]{gloudemans2023}. 
The VLA 2021 (2022) and NOEMA (ALMA) data were observed just 6.5 (2.8) days apart in the quasar's rest frame. We, therefore, interpret the differences measured at $\sim$240\,GHz to be due to synchrotron variability. Indeed, to be able to measure this variability, the synchrotron radiation must be the dominant source in the rest-frame FIR of \blazar. 

Extrapolating the measured synchrotron power law to rest-frame IR we would have expected an even brighter source (by a factor $2-3$) in the continuum detections (see Fig.~\ref{fig:sed}). 
To explain the spectral energy distribution (SED), the synchrotron emission must have a break, steepening at frequencies in between our cm and mm measurements. 
Such a break is expected for young radio sources given that higher-energy electrons radiate their energy faster and therefore have shorter lifetimes \citep[e.g.,][]{carilli1991}. 
Similar SED behavior has been observed in other powerful radio emitting $z\gtrsim 6$ quasars, and it has been used as an indirect argument that the synchrotron emission can affect the inferred FIR properties of the quasar host galaxies \citep{rojas2021, khusanova2022}.

In the case of \blazar, we have direct evidence of the synchrotron emission affecting the IR properties. If we were to assume that the NOEMA/ALMA data trace cold dust from the host galaxies---as in most studies \citep[e.g.,][]{venemans2020, izumi2019}---we would obtain very different IR luminosities, cold dust masses (not reported to avoid confusion), and star-formation rates from one epoch to the other (Fig.~\ref{fig:sed}). 
For example, the inferred SFR$_{\rm IR}$ following \cite{kennicutt1998}  would be $400\pm 29\,M_\odot\,$yr$^{-1}$ and $295\pm 21\,M_\odot\,$yr$^{-1}$ in 2021 and 2022, respectively. That is 4--9 times larger than what is inferred from the \cii\ line (see Table~\ref{tab:properties}). The dust masses and star-formation rates do not change in a month. Thus, one must be careful of blindly interpreting rest-frame IR continuum emission as fully originating from dust, at least for quasars with evidence of relativistic jets.

\subsection{[CII] deficit}
\label{sec:ciideficit}

The observed trend of decreasing \cii\ to total IR luminosity for galaxies with $L_{\rm IR}> 10^{11}\,M_\odot$ is known as the ``\cii\ deficit'' \citep[e.g.,][]{malhotra2001, diaz-santos2013}. The physical explanation for this trend is still an open question and possible explanations include the effects of AGN, the intensity of the radiation field, and optically thick \cii\ emission \citep[e.g.,][]{casey2014,lagache2018}. 

\begin{figure*}[h!]
\centering
\includegraphics[width=0.95\linewidth]{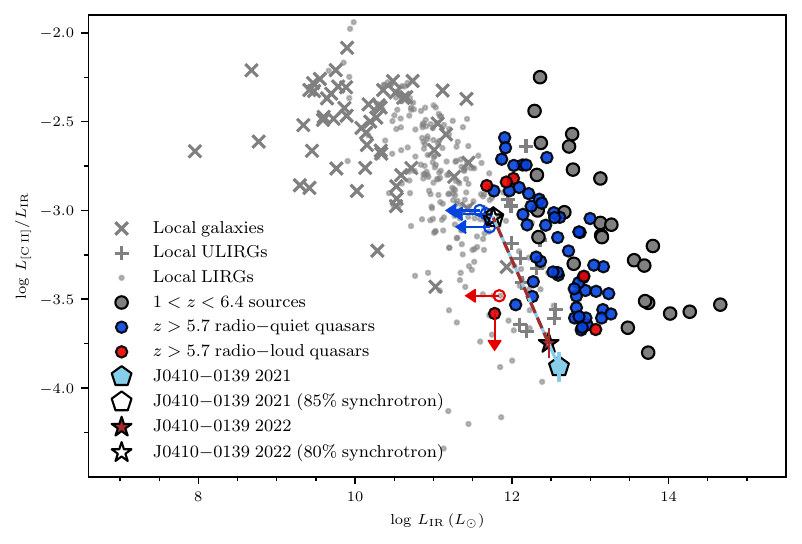}
\caption{
\textbf{Ratio of [CII] to IR luminosities ($8-1000\,\mu$m) vs.\ IR luminosities.  } The local galaxies and the high-redshift sources that are \textit{not} quasars are compiled as in \cite{banados2015b}, including data from \cite{malhotra2001, luhman2003, diaz-santos2013,delooze2014}. The $z>5.7$ radio-quiet quasars are from \cite{venemans2020} and Wang et al.\ (2024). The  FIR ($42.5-122.5\,\mu$m) luminosities from \cite{venemans2020} are converted as $L_{\rm TIR}= 1.41\times L_{\rm FIR}$. 
The $z>5.7$ radio-loud quasars are from \cite{khusanova2022} and \cite{banados2023}. The open circles are sources undetected in both \cii\ and IR continuum. 
The red star and blue pentagon correspond to the properties of \blazar\ using the IR properties derived from NOEMA in 2021 and ALMA in 2022, respectively. At face value, these measurements would place \blazar\ as the lowest data point in $L_{\rm \cii}/L_{\rm IR}$ among the $z>5.7$ quasars. However, we argue that the IR luminosity is contaminated by synchrotron emission, as evidenced by variability. The open star and pentagons show the location of \blazar, consistent with local (U)LIRGs, if we require that the SFR derived from IR luminosity matches the one derived from \cii\ line. In this case, the measured IR luminosity is dominated by synchrotron radiation: $\sim 85\%$ in 2021 and $\sim 80\%$ in 2022. 
}
\label{fig:ciidefilcit}
\end{figure*}

In general, the $z>5.7$ quasars follow the \cii\ deficit as shown in Figure~\ref{fig:ciidefilcit}. Based on its measured IR flux density, \blazar\ would be the $z>5.7$ quasar with the lowest \cii-to-IR luminosity ratio reported to date, ranging from  $\log  L_{\rm \cii}/L_{\rm IR} = -3.7$ to  $\log  L_{\rm \cii}/L_{\rm IR} = -4.0$ (see Figure~\ref{fig:ciidefilcit}). From Section \ref{sec:variability}, we know that \blazar's IR luminosity is not only due to cold dust but is also affected---and most likely dominated---by synchrotron emission. We can obtain an order-of-magnitude estimate of the synchrotron contribution to the IR luminosity, considering that the host galaxy contribution is not expected to vary in scales of months.  
Under the assumption that dust-obscured star formation dominates for massive galaxies \cite[e.g.,][]{ferrara2022,algera2023}, to make the \cii- and IR-derived SFR compatible, the synchrotron contribution needs to be 85\% and 80\% in the mm observations from 2021 and 2022, respectively. Taking this into account, the host galaxy of \blazar\ would comfortably be located in the expected region for local (U)LIRG galaxies in the $L_{\rm \cii}/L_{\rm IR}$ vs.\ $L_{\rm \cii}$ plane (Fig.~\ref{fig:ciidefilcit}). 

We can speculate that some synchrotron contribution to the IR luminosities could explain the ``\cii\ deficit'' for some of the luminous quasars. However, if we separate the radio-loud quasars (and, therefore, the sources that are most likely affected by synchrotron radiation) in Figure~\ref{fig:ciidefilcit}, the picture is not that clear. The majority of the radio-loud quasars (although the statistics are still limited) do not seem to have particularly anomalous $L_{\rm \cii}/L_{\rm IR}$.  The fraction of sources non-detected in \cii\ and/or underlying continuum seems to be higher for radio-loud quasars. However, the current number statistics are limited, and therefore,  a larger sample of radio-loud quasars with \cii\ observations is required to robustly determine any potential differences with the measurements on radio-quiet quasars.

\subsection{Nearby continuum source}
There is a continuum source in the ALMA data at R.A$=$04:10:10.01 and Decl.$=-$01:39.19.40 ($14\farcs37$ from \blazar). 
The continuum emission is marginally resolved, with a deconvolved size of $(0.66\pm 0.26)^{\prime\prime}\times (0.43\pm 0.26)^{\prime\prime}$. 
The integrated flux density is $S_{238\,\mathrm{GHz}}=0.20 \pm 0.04$\,mJy. No line is detected in a spectrum extracted from the brightest pixel or a $1\farcs5$ radius aperture. The source is undetected in all the other existing multi-wavelength datasets in this field (see \citealt{banados2024a}); thus, its nature and redshift remain unknown.

\begin{figure*}[h!]
\centering
\includegraphics[width=0.95\linewidth]{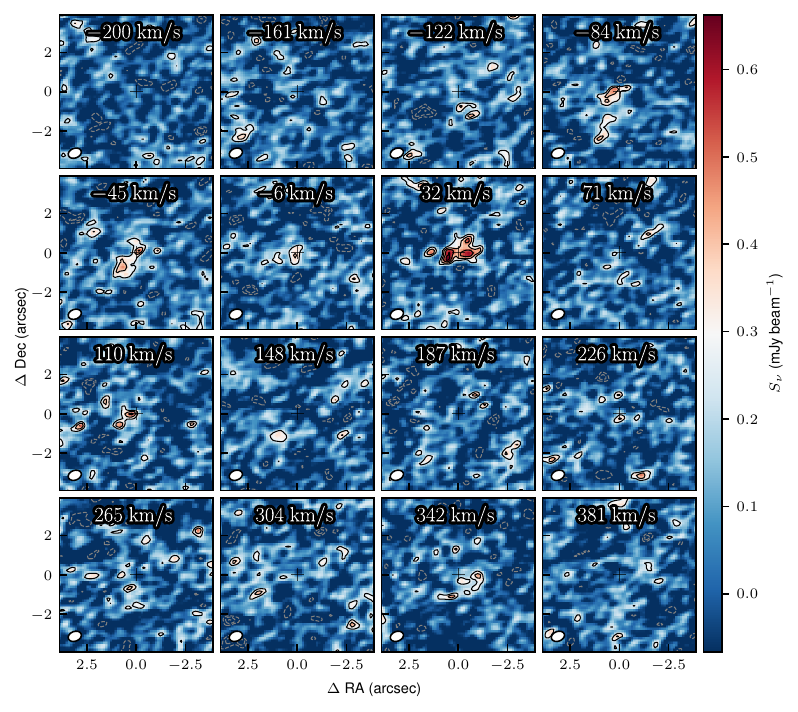}
\caption{\small
\textbf{40\,\kms\ [CII] channel maps of the continuum subtracted data of \blazar.} Solid (dashed) contours correspond to positive (negative) values starting at $2\sigma$ with steps of $1\sigma$, where $\sigma$ is the noise in each individual channel. The synthesized beam is shown in the lower-left corner of each panel. 
}
\label{fig:chmaps30}
\end{figure*}

\begin{figure*}[h!]
\centering
\includegraphics[width=0.95\linewidth]{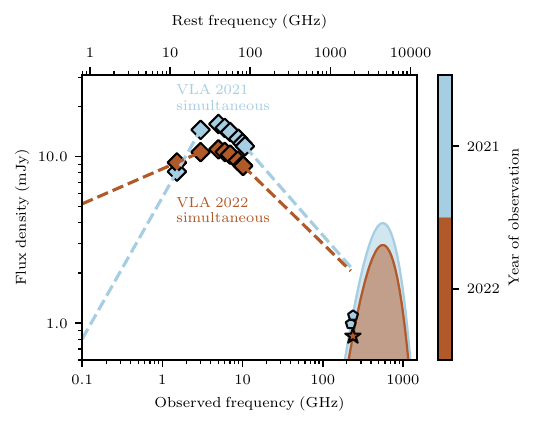}
\caption{\small
\textbf{Spectral energy distribution (SED) in the radio and (sub)mm regime of \blazar}. The data points are color-coded by year of observation for visualization purposes. Only the observations taken at a similar epoch in 2021 and 2022 are shown (see Fig.~2 in \citealt{banados2024a} for the full SED and other epochs). The VLA 2021 and 2022 observations were taken in the same observing block, yielding a quasi-simultaneous view of the SED  (within 15 minutes in the rest frame). 
The VLA 2021 and NOEMA observations are taken 6.5 days apart in the source's rest frame, while the VLA 2022 and ALMA observations are 2.8 days apart. The dashed lines are power-law extrapolations from the VLA low and high-frequency measurements. The pentagons and star show the NOEMA and ALMA measurements, respectively (see also Fig.~\ref{fig:alma_noema}). The $\sim 238\,$GHz continuum decreased by $\sim$33\% from the NOEMA 2021 to ALMA 2022 measurements. The VLA flux densities above 3\,GHz decreased on average by $\sim 37\%$ from 2021 to 2022 (see Sect.~\ref{sec:variability}), strongly suggesting that the differences between the NOEMA and ALMA data are due to the variability of the synchrotron emission. 
In contrast, we show models of modified black-body emission (with typical dust temperatures of $T_{\rm d}=47\,$K and emissivity index of $\beta=1.6$) with the colored curves. By failing to account for the IR contribution of the synchrotron emission, single-epoch measurements of cold gas in the host galaxy would be very different---and incorrect.
}
\label{fig:sed}
\end{figure*}

\section{Summary and concluding remarks}

Here, we presented NOEMA and ALMA observations of the highest redshift blazar currently known, \blazar. 
The main results can be summarized as:

\begin{itemize}
\item[(i)] The \cii\ emission line is detected in the ALMA data (Figure~\ref{fig:alma_noema}), providing a systemic redshift of $z_{\cii} = 6.9964\pm 0.0005$, in good agreement with the redshift derived from the quasar's broad line region ($z_{\mgii}=6.995 \pm 0.001$ ; $\Delta v = 53 \pm 42\,\kms$; \citealt{banados2024a}). The \cii\ measurements and derived properties are listed in Table~\ref{tab:properties}. 

\item[(ii)] The centroid of the \cii\ emission is offset by 0\farcs35 (1.81\,kpc) from the position of the quasar (Figure~\ref{fig:alma_noema}). The \cii\ emission reveals multiple peaks and different substructures in separate channel maps (Figure~\ref{fig:chmaps30}). 
This could be evidence of an ongoing merger; however, with the current sensitivity of the observations, it is difficult to conclusively constrain the kinematics. 

\item[(iii)] We estimate a dynamical mass of $M_{\rm dyn,\cii}=(4.6\pm 2.0)\times 10^9\,M_\odot$. We obtain black hole mass to host a dynamical mass ratio of $0.15^{+0.08}_{-0.05}$, placing the quasar above the local relation and consistent with other $z\gtrsim 6$ quasars.

\item[(iv)] We detect a decrease of $\sim33\%$ in the rest-frame IR continuum in about one month in the quasar's rest-frame (Table~\ref{tab:properties}). This decrease is strongly correlated with the variability measured at GHz frequencies by the VLA on a similar timeframe (Figure~\ref{fig:sed}). At $z<1$, contamination of synchrotron to rest-frame IR continuum emission is frequently observed in radio-loud AGN \citep[e.g.,][]{weiss2008,dicken2023}. We interpret \blazar's variability in the IR as due to synchrotron emission, providing the first direct evidence of synchrotron radiation affecting the measured IR properties of $z\gtrsim 6$ quasars. 

\item[(v)] \blazar\ shows the lowest  $L_{\rm \cii}/L_{\rm IR}$ among all $z>5.7$ quasars studied to date. We argue that this is because the synchrotron emission is dominating the IR luminosity ($\gtrsim 80\%$). The influence of synchrotron emission can help explain some of the sources showing the most extreme ``\cii\ deficit'' (Section~\ref{sec:ciideficit}). However, most of the radio-loud quasars studied to date are not extreme sources in the $L_{\rm \cii}/L_{\rm IR}$ vs.\ $L_{\rm \cii}$ plane, complicating this interpretation (Figure~\ref{fig:sed}).
\end{itemize}

This work highlights the importance of having a well-sampled and, if possible, quasi-simultaneous SED for interpreting the properties of quasars with strong radio emission.

\begin{table}[h!]
\small
\centering
\caption{\textbf{Measured and derived properties of \blazar.}}
\begin{tabular}{lll}
\hline 
Quantity  & Value & Units \\
\hline
\vspace{2pt}
$z_{\rm \mgii}$     & $6.995\pm 0.001$         &          \\
$z_{\rm \cii}$      & $6.9964\pm 0.0005$         &          \\
FWHM$_{\rm \cii}$      & $206 \pm 40$        & \kms      \\
I$_{\rm \cii}$      & $0.45 \pm 0.08$        & Jy\,\kms      \\
$L_{\rm \cii}$      & $(2.02\pm 0.36)\times 10^{42}$         &  erg\,s$^{-1}$        \\
SFR$_{\rm \cii}^a$      & $58\pm 9$         &   $M_\odot\,$yr$^{-1}$      \\
 $M_{\rm dyn,\cii}^b$ &$(4.6\pm 2.0)\times 10^9$ &$M_\odot$\\
S$_{\rm 224\,GHz}$ (NOEMA2021)      & $0.99\pm 0.07$         &   mJy     \\
S$_{\rm 239\,GHz}$ (NOEMA2021)      & $1.11\pm 0.09$         &   mJy     \\
S$_{\rm 238\,GHz}$ (ALMA2022)      & $0.84\pm 0.06$          &   mJy     \\
$L_{\rm IR}(2021)^c$      & $(1.53\pm 0.11)\times 10^{46}$         &  erg\,s$^{-1}$        \\
$L_{\rm IR}(2022)^c$      & $(1.13\pm 0.08)\times 10^{46}$         &  erg\,s$^{-1}$        \\
\hline
\label{tab:properties}
\end{tabular}
\\
\footnotesize{
$^a$ The SFR$_{\rm \cii}$ is calculated using $\log \mathrm{SFR}_{\rm \cii}=-6.09+0.90\times \log L_{\rm \cii}$, with $L_{\rm \cii}$ in solar luminosities \citep{delooze2014}. \\
$^b$ See Sect.~\ref{sec:ciiprop} for the dynamical mass estimation.\\
$^c$ The IR luminosities are integrated over $8-1000\,\mu$m assuming a modified black body with dust temperature of $T_d=47\,K$ and emissivity index $\beta=1.6$ (see Fig.~\ref{fig:sed}). 
} 
\end{table}

\begin{acknowledgments}
We thank the staff of IRAM/NOEMA and ALMA for making these observations possible. 
YK thanks the support of the German Space Agency (DLR) through the program LEGACY 50OR2303. 
CM acknowledges support from Fondecyt Iniciacion grant 11240336  and ANID BASAL project FB210003.   
Based on observations carried out with the IRAM Interferometer NOEMA (program S21DK). IRAM is supported by INSU/CNRS (France), MPG (Germany) and IGN (Spain). 
This paper makes use of the following ALMA data: ADS/JAO.ALMA\#2021.A.00030.S. ALMA is a partnership of ESO (representing its member states), NSF (USA) and NINS (Japan), together with NRC (Canada), NSTC and ASIAA (Taiwan), and KASI (Republic of Korea), in cooperation with the Republic of Chile. The Joint ALMA Observatory is operated by ESO, AUI/NRAO and NAOJ.
\end{acknowledgments}

%

\vspace{5mm}
\facilities{NOEMA, ALMA}


\software{astropy \citep{2013A&A...558A..33A,2018AJ....156..123A},
CASA \citep{mcmullin2007},
GILDAS \url{https://www.iram.fr/IRAMFR/GILDAS},
interferopy \cite{interferopy},
Matplotlib \citep[][\url{http://www.matplotlib.org}]{hunter2007}
          }




\bibliography{library_nameyyyy}{}

\begin{thebibliography}{}
\expandafter\ifx\csname natexlab\endcsname\relax\def\natexlab#1{#1}\fi
\providecommand{\url}[1]{\href{#1}{#1}}
\providecommand{\dodoi}[1]{doi:~\href{http://doi.org/#1}{\nolinkurl{#1}}}
\providecommand{\doeprint}[1]{\href{http://ascl.net/#1}{\nolinkurl{http://ascl.net/#1}}}
\providecommand{\doarXiv}[1]{\href{https://arxiv.org/abs/#1}{\nolinkurl{https://arxiv.org/abs/#1}}}

\bibitem[{{Algera} {et~al.}(2023){Algera}, {Inami}, {Oesch}, {Sommovigo}, {Bouwens}, {Topping}, {Schouws}, {Stefanon}, {Stark}, {Aravena}, {Barrufet}, {da Cunha}, {Dayal}, {Endsley}, {Ferrara}, {Fudamoto}, {Gonzalez}, {Graziani}, {Hodge}, {Hygate}, {de Looze}, {Nanayakkara}, {Schneider}, \& {van der Werf}}]{algera2023}
{Algera}, H. S.~B., {Inami}, H., {Oesch}, P.~A., {et~al.} 2023, \mnras, 518, 6142, \dodoi{10.1093/mnras/stac3195}

\bibitem[{{Astropy Collaboration} {et~al.}(2013){Astropy Collaboration}, {Robitaille}, {Tollerud}, {Greenfield}, {Droettboom}, {Bray}, {Aldcroft}, {Davis}, {Ginsburg}, {Price-Whelan}, {Kerzendorf}, {Conley}, {Crighton}, {Barbary}, {Muna}, {Ferguson}, {Grollier}, {Parikh}, {Nair}, {Unther}, {Deil}, {Woillez}, {Conseil}, {Kramer}, {Turner}, {Singer}, {Fox}, {Weaver}, {Zabalza}, {Edwards}, {Azalee Bostroem}, {Burke}, {Casey}, {Crawford}, {Dencheva}, {Ely}, {Jenness}, {Labrie}, {Lim}, {Pierfederici}, {Pontzen}, {Ptak}, {Refsdal}, {Servillat}, \& {Streicher}}]{2013A&A...558A..33A}
{Astropy Collaboration}, {Robitaille}, T.~P., {Tollerud}, E.~J., {et~al.} 2013, \aap, 558, A33, \dodoi{10.1051/0004-6361/201322068}

\bibitem[{{Astropy Collaboration} {et~al.}(2018){Astropy Collaboration}, {Price-Whelan}, {Sip{\H{o}}cz}, {G{\"u}nther}, {Lim}, {Crawford}, {Conseil}, {Shupe}, {Craig}, {Dencheva}, {Ginsburg}, {Vand erPlas}, {Bradley}, {P{\'e}rez-Su{\'a}rez}, {de Val-Borro}, {Aldcroft}, {Cruz}, {Robitaille}, {Tollerud}, {Ardelean}, {Babej}, {Bach}, {Bachetti}, {Bakanov}, {Bamford}, {Barentsen}, {Barmby}, {Baumbach}, {Berry}, {Biscani}, {Boquien}, {Bostroem}, {Bouma}, {Brammer}, {Bray}, {Breytenbach}, {Buddelmeijer}, {Burke}, {Calderone}, {Cano Rodr{\'\i}guez}, {Cara}, {Cardoso}, {Cheedella}, {Copin}, {Corrales}, {Crichton}, {D'Avella}, {Deil}, {Depagne}, {Dietrich}, {Donath}, {Droettboom}, {Earl}, {Erben}, {Fabbro}, {Ferreira}, {Finethy}, {Fox}, {Garrison}, {Gibbons}, {Goldstein}, {Gommers}, {Greco}, {Greenfield}, {Groener}, {Grollier}, {Hagen}, {Hirst}, {Homeier}, {Horton}, {Hosseinzadeh}, {Hu}, {Hunkeler}, {Ivezi{\'c}}, {Jain}, {Jenness}, {Kanarek}, {Kendrew}, {Kern}, {Kerzendorf}, {Khvalko}, {King}, {Kirkby}, {Kulkarni},
  {Kumar}, {Lee}, {Lenz}, {Littlefair}, {Ma}, {Macleod}, {Mastropietro}, {McCully}, {Montagnac}, {Morris}, {Mueller}, {Mumford}, {Muna}, {Murphy}, {Nelson}, {Nguyen}, {Ninan}, {N{\"o}the}, {Ogaz}, {Oh}, {Parejko}, {Parley}, {Pascual}, {Patil}, {Patil}, {Plunkett}, {Prochaska}, {Rastogi}, {Reddy Janga}, {Sabater}, {Sakurikar}, {Seifert}, {Sherbert}, {Sherwood-Taylor}, {Shih}, {Sick}, {Silbiger}, {Singanamalla}, {Singer}, {Sladen}, {Sooley}, {Sornarajah}, {Streicher}, {Teuben}, {Thomas}, {Tremblay}, {Turner}, {Terr{\'o}n}, {van Kerkwijk}, {de la Vega}, {Watkins}, {Weaver}, {Whitmore}, {Woillez}, {Zabalza}, \& {Astropy Contributors}}]{2018AJ....156..123A}
{Astropy Collaboration}, {Price-Whelan}, A.~M., {Sip{\H{o}}cz}, B.~M., {et~al.} 2018, \aj, 156, 123, \dodoi{10.3847/1538-3881/aabc4f}

\bibitem[{{Ba{\~n}ados} {et~al.}(2015){Ba{\~n}ados}, {Decarli}, {Walter}, {Venemans}, {Farina}, \& {Fan}}]{banados2015b}
{Ba{\~n}ados}, E., {Decarli}, R., {Walter}, F., {et~al.} 2015, \apjl, 805, L8, \dodoi{10.1088/2041-8205/805/1/L8}

\bibitem[{{Ba{\~n}ados} {et~al.}(2019){Ba{\~n}ados}, {Novak}, {Neeleman}, {Walter}, {Decarli}, {Venemans}, {Mazzucchelli}, {Carilli}, {Wang}, {Fan}, {Farina}, \& {Rix}}]{banados2019a}
{Ba{\~n}ados}, E., {Novak}, M., {Neeleman}, M., {et~al.} 2019, \apjl, 881, L23, \dodoi{10.3847/2041-8213/ab3659}

\bibitem[{{Ba{\~n}ados} {et~al.}(2023){Ba{\~n}ados}, {Schindler}, {Venemans}, {Connor}, {Decarli}, {Farina}, {Mazzucchelli}, {Meyer}, {Stern}, {Walter}, {Fan}, {Hennawi}, {Khusanova}, {Morrell}, {Nanni}, {Noirot}, {Pensabene}, {Rix}, {Simon}, {Verdoes Kleijn}, {Xie}, {Yang}, \& {Connor}}]{banados2023}
{Ba{\~n}ados}, E., {Schindler}, J.-T., {Venemans}, B.~P., {et~al.} 2023, \apjs, 265, 29, \dodoi{10.3847/1538-4365/acb3c7}

\bibitem[{{Ba{\~n}ados} {et~al.}(2024){Ba{\~n}ados}, {Momjian}, {Connor}, {Belladitta}, {Decarli}, {Mazzucchelli}, {Venemans}, {Walter}, {Wang}, {Xie}, {Barth}, {Eilers}, {Fan}, {Khusanova}, {Schindler}, {Stern}, {Yang}, {Taufik Andika}, {Carilli}, {Farina}, {Fabian}, {Hennawi}, {Pensabene}, \& {Rojas-Ruiz}}]{banados2024a}
{Ba{\~n}ados}, E., {Momjian}, E., {Connor}, T., {et~al.} 2024, arXiv e-prints, arXiv:2407.07236.
\newblock \doarXiv{2407.07236}

\bibitem[{{Bieri} {et~al.}(2015){Bieri}, {Dubois}, {Silk}, \& {Mamon}}]{bieri2015}
{Bieri}, R., {Dubois}, Y., {Silk}, J., \& {Mamon}, G.~A. 2015, \apjl, 812, L36, \dodoi{10.1088/2041-8205/812/2/L36}

\bibitem[{Boogaard {et~al.}(2021)Boogaard, Meyer, \& Novak}]{interferopy}
Boogaard, L., Meyer, R.~A., \& Novak, M. 2021, {Interferopy: analysing datacubes from radio-to-submm observations}, \dodoi{10.5281/ZENODO.5775603}

\bibitem[{{Carilli} {et~al.}(1991){Carilli}, {Perley}, {Dreher}, \& {Leahy}}]{carilli1991}
{Carilli}, C.~L., {Perley}, R.~A., {Dreher}, J.~W., \& {Leahy}, J.~P. 1991, \apj, 383, 554, \dodoi{10.1086/170813}

\bibitem[{{Casey} {et~al.}(2014){Casey}, {Narayanan}, \& {Cooray}}]{casey2014}
{Casey}, C.~M., {Narayanan}, D., \& {Cooray}, A. 2014, \physrep, 541, 45, \dodoi{10.1016/j.physrep.2014.02.009}

\bibitem[{{Comerford} {et~al.}(2020){Comerford}, {Negus}, {M{\"u}ller-S{\'a}nchez}, {Eracleous}, {Wylezalek}, {Storchi-Bergmann}, {Greene}, {Barrows}, {Nevin}, {Roy}, \& {Stemo}}]{comerford2020}
{Comerford}, J.~M., {Negus}, J., {M{\"u}ller-S{\'a}nchez}, F., {et~al.} 2020, \apj, 901, 159, \dodoi{10.3847/1538-4357/abb2ae}

\bibitem[{{Connor} {et~al.}(2024){Connor}, {Ba{\~n}ados}, {Cappelluti}, \& {Foord}}]{connor2024}
{Connor}, T., {Ba{\~n}ados}, E., {Cappelluti}, N., \& {Foord}, A. 2024, Universe, 10, 227, \dodoi{10.3390/universe10050227}

\bibitem[{{De Looze} {et~al.}(2014){De Looze}, {Cormier}, {Lebouteiller}, {Madden}, {Baes}, {Bendo}, {Boquien}, {Boselli}, {Clements}, {Cortese}, {Cooray}, {Galametz}, {Galliano}, {Graci{\'a}-Carpio}, {Isaak}, {Karczewski}, {Parkin}, {Pellegrini}, {R{\'e}my-Ruyer}, {Spinoglio}, {Smith}, \& {Sturm}}]{delooze2014}
{De Looze}, I., {Cormier}, D., {Lebouteiller}, V., {et~al.} 2014, \aap, 568, A62, \dodoi{10.1051/0004-6361/201322489}

\bibitem[{{Decarli} {et~al.}(2017){Decarli}, {Walter}, {Venemans}, {Ba{\~n}ados}, {Bertoldi}, {Carilli}, {Fan}, {Farina}, {Mazzucchelli}, {Riechers}, {Rix}, {Strauss}, {Wang}, \& {Yang}}]{decarli2017}
{Decarli}, R., {Walter}, F., {Venemans}, B.~P., {et~al.} 2017, \nat, 545, 457, \dodoi{10.1038/nature22358}

\bibitem[{{Decarli} {et~al.}(2018){Decarli}, {Walter}, {Venemans}, {Ba{\~n}ados}, {Bertoldi}, {Carilli}, {Fan}, {Farina}, {Mazzucchelli}, {Riechers}, {Rix}, {Strauss}, {Wang}, \& {Yang}}]{decarli2018}
---. 2018, \apj, 854, 97, \dodoi{10.3847/1538-4357/aaa5aa}

\bibitem[{{Decarli} {et~al.}(2019){Decarli}, {Dotti}, {Ba{\~n}ados}, {Farina}, {Walter}, {Carilli}, {Fan}, {Mazzucchelli}, {Neeleman}, {Novak}, {Riechers}, {Strauss}, {Venemans}, {Yang}, \& {Wang}}]{decarli2019}
{Decarli}, R., {Dotti}, M., {Ba{\~n}ados}, E., {et~al.} 2019, \apj, 880, 157, \dodoi{10.3847/1538-4357/ab297f}

\bibitem[{{D{\'{\i}}az-Santos} {et~al.}(2013){D{\'{\i}}az-Santos}, {Armus}, {Charmandaris}, {Stierwalt}, {Murphy}, {Haan}, {Inami}, {Malhotra}, {Meijerink}, {Stacey}, {Petric}, {Evans}, {Veilleux}, {van der Werf}, {Lord}, {Lu}, {Howell}, {Appleton}, {Mazzarella}, {Surace}, {Xu}, {Schulz}, {Sanders}, {Bridge}, {Chan}, {Frayer}, {Iwasawa}, {Melbourne}, \& {Sturm}}]{diaz-santos2013}
{D{\'{\i}}az-Santos}, T., {Armus}, L., {Charmandaris}, V., {et~al.} 2013, \apj, 774, 68, \dodoi{10.1088/0004-637X/774/1/68}

\bibitem[{{Dicken} {et~al.}(2023){Dicken}, {Tadhunter}, {Nesvadba}, {Bernhard}, {K{\"o}nyves}, {Morganti}, {Ramos Almeida}, \& {Oosterloo}}]{dicken2023}
{Dicken}, D., {Tadhunter}, C.~N., {Nesvadba}, N.~P.~H., {et~al.} 2023, \mnras, 519, 5807, \dodoi{10.1093/mnras/stac3465}

\bibitem[{{Ferrara} {et~al.}(2022){Ferrara}, {Sommovigo}, {Dayal}, {Pallottini}, {Bouwens}, {Gonzalez}, {Inami}, {Smit}, {Bowler}, {Endsley}, {Oesch}, {Schouws}, {Stark}, {Stefanon}, {Aravena}, {da Cunha}, {De Looze}, {Fudamoto}, {Graziani}, {Hodge}, {Riechers}, {Schneider}, {Algera}, {Barrufet}, {Hygate}, {Labb{\'e}}, {Li}, {Nanayakkara}, {Topping}, \& {van der Werf}}]{ferrara2022}
{Ferrara}, A., {Sommovigo}, L., {Dayal}, P., {et~al.} 2022, \mnras, 512, 58, \dodoi{10.1093/mnras/stac460}

\bibitem[{{Gloudemans} {et~al.}(2023){Gloudemans}, {Saxena}, {Intema}, {Callingham}, {Duncan}, {R{\"o}ttgering}, {Belladitta}, {Hardcastle}, {Harikane}, \& {Spingola}}]{gloudemans2023}
{Gloudemans}, A.~J., {Saxena}, A., {Intema}, H., {et~al.} 2023, \aap, 678, A128, \dodoi{10.1051/0004-6361/202347582}

\bibitem[{{Habouzit} {et~al.}(2022){Habouzit}, {Onoue}, {Ba{\~n}ados}, {Neeleman}, {Angl{\'e}s-Alc{\'a}zar}, {Walter}, {Pillepich}, {Dav{\'e}}, {Jahnke}, \& {Dubois}}]{habouzit2022}
{Habouzit}, M., {Onoue}, M., {Ba{\~n}ados}, E., {et~al.} 2022, \mnras, 511, 3751, \dodoi{10.1093/mnras/stac225}

\bibitem[{{Hardcastle} \& {Croston}(2020)}]{hardcastle2020}
{Hardcastle}, M.~J., \& {Croston}, J.~H. 2020, \nar, 88, 101539, \dodoi{10.1016/j.newar.2020.101539}

\bibitem[{{Harrison} \& {Ramos Almeida}(2024)}]{harrison2024}
{Harrison}, C.~M., \& {Ramos Almeida}, C. 2024, Galaxies, 12, 17, \dodoi{10.3390/galaxies12020017}

\bibitem[{{Herrera-Camus} {et~al.}(2015){Herrera-Camus}, {Bolatto}, {Wolfire}, {Smith}, {Croxall}, {Kennicutt}, {Calzetti}, {Helou}, {Walter}, {Leroy}, {Draine}, {Brandl}, {Armus}, {Sandstrom}, {Dale}, {Aniano}, {Meidt}, {Boquien}, {Hunt}, {Galametz}, {Tabatabaei}, {Murphy}, {Appleton}, {Roussel}, {Engelbracht}, \& {Beirao}}]{herrera-camus2015}
{Herrera-Camus}, R., {Bolatto}, A.~D., {Wolfire}, M.~G., {et~al.} 2015, \apj, 800, 1, \dodoi{10.1088/0004-637X/800/1/1}

\bibitem[{{Hunter}(2007)}]{hunter2007}
{Hunter}, J.~D. 2007, Computing in Science and Engineering, 9, 90, \dodoi{10.1109/MCSE.2007.55}

\bibitem[{{Izumi} {et~al.}(2019){Izumi}, {Onoue}, {Matsuoka}, {Nagao}, {Strauss}, {Imanishi}, {Kashikawa}, {Fujimoto}, {Kohno}, {Toba}, {Umehata}, {Goto}, {Ueda}, {Shirakata}, {Silverman}, {Greene}, {Harikane}, {Hashimoto}, {Ikarashi}, {Iono}, {Iwasawa}, {Lee}, {Minezaki}, {Nakanishi}, {Tamura}, {Tang}, \& {Taniguchi}}]{izumi2019}
{Izumi}, T., {Onoue}, M., {Matsuoka}, Y., {et~al.} 2019, \pasj, 71, 111, \dodoi{10.1093/pasj/psz096}

\bibitem[{{Keller} {et~al.}(2024){Keller}, {Thyagarajan}, {Kumar}, {Kanekar}, \& {Bernardi}}]{keller2024}
{Keller}, P.~M., {Thyagarajan}, N., {Kumar}, A., {Kanekar}, N., \& {Bernardi}, G. 2024, \mnras, 528, 5692, \dodoi{10.1093/mnras/stae418}

\bibitem[{{Kennicutt}(1998)}]{kennicutt1998}
{Kennicutt}, Jr., R.~C. 1998, \apj, 498, 541, \dodoi{10.1086/305588}

\bibitem[{{Khusanova} {et~al.}(2022){Khusanova}, {Ba{\~n}ados}, {Mazzucchelli}, {Rojas-Ruiz}, {Momjian}, {Walter}, {Decarli}, {Venemans}, {Farina}, {Meyer}, {Wang}, \& {Yang}}]{khusanova2022}
{Khusanova}, Y., {Ba{\~n}ados}, E., {Mazzucchelli}, C., {et~al.} 2022, \aap, 664, A39, \dodoi{10.1051/0004-6361/202243660}

\bibitem[{{Kormendy} \& {Ho}(2013)}]{kormendy2013}
{Kormendy}, J., \& {Ho}, L.~C. 2013, \araa, 51, 511, \dodoi{10.1146/annurev-astro-082708-101811}

\bibitem[{{Lagache} {et~al.}(2018){Lagache}, {Cousin}, \& {Chatzikos}}]{lagache2018}
{Lagache}, G., {Cousin}, M., \& {Chatzikos}, M. 2018, \aap, 609, A130, \dodoi{10.1051/0004-6361/201732019}

\bibitem[{{Li} {et~al.}(2024){Li}, {Wang}, {Pensabene}, {Walter}, {Venemans}, {Decarli}, {Ba{\~n}ados}, {Cox}, {Neri}, {Omont}, {Cai}, {Khusanova}, {Xu}, {Riechers}, {Wagg}, {Shao}, {Liu}, {Menten}, {Li}, \& {Fan}}]{li2024}
{Li}, J., {Wang}, R., {Pensabene}, A., {et~al.} 2024, \apj, 962, 119, \dodoi{10.3847/1538-4357/ad1754}

\bibitem[{{Luhman} {et~al.}(2003){Luhman}, {Satyapal}, {Fischer}, {Wolfire}, {Sturm}, {Dudley}, {Lutz}, \& {Genzel}}]{luhman2003}
{Luhman}, M.~L., {Satyapal}, S., {Fischer}, J., {et~al.} 2003, \apj, 594, 758, \dodoi{10.1086/376965}

\bibitem[{{Malhotra} {et~al.}(2001){Malhotra}, {Kaufman}, {Hollenbach}, {Helou}, {Rubin}, {Brauher}, {Dale}, {Lu}, {Lord}, {Stacey}, {Contursi}, {Hunter}, \& {Dinerstein}}]{malhotra2001}
{Malhotra}, S., {Kaufman}, M.~J., {Hollenbach}, D., {et~al.} 2001, \apj, 561, 766, \dodoi{10.1086/323046}

\bibitem[{{Marshall} {et~al.}(2023){Marshall}, {Perna}, {Willott}, {Maiolino}, {Scholtz}, {{\"U}bler}, {Carniani}, {Arribas}, {L{\"u}tzgendorf}, {Bunker}, {Charlot}, {Ferruit}, {Jakobsen}, {Rix}, {Rodr{\'\i}guez Del Pino}, {B{\"o}ker}, {Cameron}, {Cresci}, {Curtis-Lake}, {Jones}, {Kumari}, {P{\'e}rez-Gonz{\'a}lez}, \& {Reed}}]{marshall2023}
{Marshall}, M.~A., {Perna}, M., {Willott}, C.~J., {et~al.} 2023, \aap, 678, A191, \dodoi{10.1051/0004-6361/202346113}

\bibitem[{{McMullin} {et~al.}(2007){McMullin}, {Waters}, {Schiebel}, {Young}, \& {Golap}}]{mcmullin2007}
{McMullin}, J.~P., {Waters}, B., {Schiebel}, D., {Young}, W., \& {Golap}, K. 2007, in Astronomical Society of the Pacific Conference Series, Vol. 376, Astronomical Data Analysis Software and Systems XVI, ed. R.~A. {Shaw}, F.~{Hill}, \& D.~J. {Bell}, 127

\bibitem[{Neeleman {et~al.}(2019)Neeleman, Ba{\~{n}}ados, Walter, Decarli, Venemans, Carilli, Fan, Farina, Mazzucchelli, Novak, Riechers, Rix, \& Wang}]{neeleman2019}
Neeleman, M., Ba{\~{n}}ados, E., Walter, F., {et~al.} 2019, \apj, 882, 10, \dodoi{10.3847/1538-4357/ab2ed3}

\bibitem[{{Neeleman} {et~al.}(2021){Neeleman}, {Novak}, {Venemans}, {Walter}, {Decarli}, {Kaasinen}, {Schindler}, {Ba{\~n}ados}, {Carilli}, {Drake}, {Fan}, \& {Rix}}]{neeleman2021}
{Neeleman}, M., {Novak}, M., {Venemans}, B.~P., {et~al.} 2021, \apj, 911, 141, \dodoi{10.3847/1538-4357/abe70f}

\bibitem[{{Novak} {et~al.}(2020){Novak}, {Venemans}, {Walter}, {Neeleman}, {Kaasinen}, {Liang}, {Feldmann}, {Ba{\~n}ados}, {Carilli}, {Decarli}, {Drake}, {Fan}, {Farina}, {Mazzucchelli}, {Rix}, \& {Wang}}]{novak2020}
{Novak}, M., {Venemans}, B.~P., {Walter}, F., {et~al.} 2020, \apj, 904, 131, \dodoi{10.3847/1538-4357/abc33f}

\bibitem[{{Rojas-Ruiz} {et~al.}(2021){Rojas-Ruiz}, {Ba{\~n}ados}, {Neeleman}, {Connor}, {Eilers}, {Venemans}, {Khusanova}, {Carilli}, {Mazzucchelli}, {Decarli}, {Momjian}, \& {Novak}}]{rojas2021}
{Rojas-Ruiz}, S., {Ba{\~n}ados}, E., {Neeleman}, M., {et~al.} 2021, \apj, 920, 150, \dodoi{10.3847/1538-4357/ac1a13}

\bibitem[{{Schreiber} {et~al.}(2015){Schreiber}, {Pannella}, {Elbaz}, {B{\'e}thermin}, {Inami}, {Dickinson}, {Magnelli}, {Wang}, {Aussel}, {Daddi}, {Juneau}, {Shu}, {Sargent}, {Buat}, {Faber}, {Ferguson}, {Giavalisco}, {Koekemoer}, {Magdis}, {Morrison}, {Papovich}, {Santini}, \& {Scott}}]{schreiber2015}
{Schreiber}, C., {Pannella}, M., {Elbaz}, D., {et~al.} 2015, \aap, 575, A74, \dodoi{10.1051/0004-6361/201425017}

\bibitem[{{Venemans} {et~al.}(2020){Venemans}, {Walter}, {Neeleman}, {Novak}, {Otter}, {Decarli}, {Ba{\~n}ados}, {Drake}, {Farina}, {Kaasinen}, {Mazzucchelli}, {Carilli}, {Fan}, {Rix}, \& {Wang}}]{venemans2020}
{Venemans}, B.~P., {Walter}, F., {Neeleman}, M., {et~al.} 2020, \apj, 904, 130, \dodoi{10.3847/1538-4357/abc563}

\bibitem[{{Wei{\ss}} {et~al.}(2008){Wei{\ss}}, {Kov{\'a}cs}, {G{\"u}sten}, {Menten}, {Schuller}, {Siringo}, \& {Kreysa}}]{weiss2008}
{Wei{\ss}}, A., {Kov{\'a}cs}, A., {G{\"u}sten}, R., {et~al.} 2008, \aap, 490, 77, \dodoi{10.1051/0004-6361:200809909}

\end{thebibliography}
\bibliographystyle{aasjournal}



\end{document}